\documentclass[12pt]{article}
\usepackage{graphicx}
\usepackage{a4}
\usepackage{amsfonts}
\usepackage{amssymb}
\usepackage{amsmath}
\usepackage{epsfig}

\begin{document}
\noindent{\large\bf

Resonances and Decay Widths within a\\ Relativistic Coupled
Channel Approach

}\vskip 4mm
%

Regina Kleinhappel, Wolfgang Schweiger

\vskip 4mm
{\small

Institut f\"ur Physik, Universit\"at Graz, A-8010 Graz, Austria
\\ (regina.kleinhappel@gmx.at, wolfgang.schweiger@uni-graz.at)

}
\vspace{4mm}


Within constituent-quark models the resonance character of hadron
excitations is usually ignored. They rather come out as stable
bound states and their bound-state wave function is then used to
calculate partial decay widths perturbatively by assuming a
particular model for the elementary decay vertex. The fact that
the predicted strong decay widths are notoriously too small
\cite{Melde:2006yw,Metsch:2008zz} is an indication that a physical
hadron resonance is not just a simple bound state of valence
(anti)quarks, but it should contain also (anti)quark-meson
components.

A good starting point to take such components into account is the
chiral  constituent quark model ($\chi
QCM$)~\cite{Glozman:1995fu}. The effective degrees of freedom of
the $\chi QCM$, that are assumed to emerge from chiral symmetry
breaking of QCD,  are constituent (anti)\-quarks and Goldstone
bosons which couple directly to the (anti)\-quarks. In order to
take relativity fully into account we work within point-form
quantum mechanics~\cite{Dirac:1949cp}, which is characterized by
the property that the components of the four momentum operator
$\hat{P}^{\mu}$ are the only generators of the Poincar\'{e} group
which contain interaction terms. A convenient method to add
interactions to $\hat{P}^{\mu}_{\mathrm{free}}$ such that the
Poincar\'{e} algebra is satisfied is the Bakamijan-Thomas
construction~\cite{Bakamjian:1953kh}. The point-form version of
the Bakamjian-Thomas construction amounts to factorize the free
4-momentum operator into a free mass operator
$\hat{M}_{\mathrm{free}}$ and a free 4-velocity operator
$V^\mu_{\mathrm{free}}$ and to add a Lorentz-scalar interaction
term $\hat{M}_{\mathrm{int}}$ that should also commute with
$\hat{V}^\mu_{\mathrm{free}}$ to $\hat{M}_{\mathrm{free}}$. The
interacting 4-momentum operator then has the structure
\begin{equation}
\hat{P}^\mu=\hat{P}^{\mu}_{\mathrm{free}}+\hat{P}^{\mu}_{\mathrm{int}}
=(\hat{M}_{\mathrm{free}}+\hat{M}_{\mathrm{int}})\,\hat{V}^\mu_{\mathrm{free}}\,
,
\end{equation}
and one only needs to study an eigenvalue problem for the mass
operator. A very useful basis, which is tailored to this kind of
construction, is formed by velocity states~\cite{Klink:1998zz}.
These are usual momentum states in the center-of-momentum of the
whole system which are then boosted to the overall four-velocity
$v^{\mu}$. In this basis the usual addition rules of
nonrelativistic quantum mechanics can be applied to spin and
angular momentum.

In order to allow for the decay of hadron excitations into a lower
lying state and a Goldstone boson we formulate the eigenvalue
problem for the mass operator as a 2-channel problem. A general
mass eigenstate is then a direct sum of a valence (anti)quark
component and a valence (antiquark) + Goldstone-boson component.
The latter can be eliminated by means of a Feshbach reduction and
one ends up with a mass-eigenvalue equation for the valence
(anti)quark component. In case of a meson, e.g., this equation
takes on the form:
\begin{equation}\label{eq:mev}
\left(\hat{M}_{q\bar{q}}+\underbrace{\hat K^\dagger (\hat
M_{q\bar{q} \pi}-m)^{-1}\hat K}_{\hat
V_{\mathrm{opt}}(m)}\right)\vert \psi_{q\bar{q}}\rangle = m \vert
\psi_{q\bar{q}}\rangle \, .
\end{equation}
The channel mass operator $\hat{M}_{q\bar{q}}$ is assumed to
contain already an instantaneous confinement and the optical
potential $V_{\mathrm{opt}}(m)$ describes all four possibilities
for the (dynamical) exchange of a Goldstone boson between
anti(quark) and (anti)quark, in particular also reabsorption of
the Goldstone boson by the emitting (anti)quark. Here we have
taken the $\pi$ as a representative for the Goldstone bosons. The
vertex operator $\hat{K}$ is derived from an appropriate field
theoretical interaction Lagrangian density~\cite{Klink:2000pp}.
\begin{figure}
\begin{center}
\includegraphics[height=3.0cm]{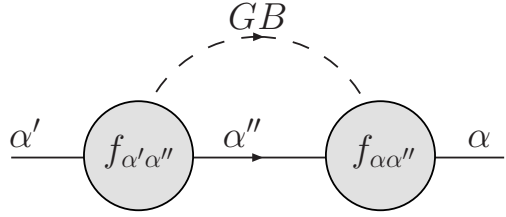}
\end{center} \caption{\small{Graphical representation of the optical potential,
$V^{nn^{\prime}}_{\mathrm{opt}}(m)$ that enters the mass
eigenvalue equation~(\ref{EVEquation}) on the hadronic level.}}
\label{fig:vopt}
\end{figure}

In a velocity state representation Eq.~(\ref{eq:mev}) becomes an
integral equation. In order to make it better amenable to a
numerical treatment we expand $\vert \psi_{q\bar{q}}\rangle$ in
terms of (velocity) eigenstates $\vert v, \alpha\rangle$ of
$\hat{M}_{q\bar{q}}$, i.e. the pure confinement problem. $\alpha$
collectively denotes the internal quantum numbers that specify
these states. For reasons which will become clear immediately, we
call $\vert v, \alpha\rangle$ a \lq\lq bare\rq\rq\ meson state,
whereas $\vert \psi_{q\bar{q}}\rangle$ is (the $q$-$\bar{q}$
component of) a \lq\lq physical\rq\rq\ meson state. This expansion
leads to an infinite set of coupled algebraic equations for the
expansion coefficients $A_\alpha$:
\begin{eqnarray}
\label{EVEquation}
&&\sum_{\alpha^{\prime}}\left(m_{\alpha}\delta_{\alpha
\alpha^{\prime}}+ V^{\alpha
\alpha^{\prime}}_{\mathrm{opt}}(m)\right) A_{\alpha^{\prime}}= m
A_{\alpha}\, .
\end{eqnarray}
The most remarkable feature of this equation is that it is rather
a mass-eigenvalue equation for mesons than for quarks. It
describes how a physical meson of mass $m$ is composed of bare
mesons with masses $m_\alpha$. The bare mesons are mixed via the
optical-potential matrix elements
$V^{nn^{\prime}}_{\mathrm{opt}}(m)$. Even these matrix elements
attain a rather simple interpretation in terms of hadronic degrees
of freedom. They couple a bare meson state with quantum numbers
$\alpha^\prime$ to another bare meson state with quantum numbers
$\alpha$ via a Goldstone-boson loop such that any bare meson state
with quantum numbers $\alpha^{\prime\prime}$ (that is allowed by
conservation laws) can be excited in an intermediate step (see
Fig.~\ref{fig:vopt}). $f_{\alpha \alpha^{\prime}}(\vert
\vec{\kappa}\vert)$ are (strong) transition form factors that show
up at the (bare) meson Goldstone-boson vertices. The eigenvalue
problem that one ends up with describes thus bare mesons, i.e.
eigenstates of the pure confinement problem, that are mixed and
dressed via Goldstone-boson loops. The only places where the quark
substructure enters, are the vertex form factors. Here it should
be emphasized that due to the instantaneous nature of the
confinement potential the dressing happens on the hadron level and
not on the quark level, i.e. emission and absorption of the
Goldstone boson by the same constituent must not be interpreted as
mass renormalization of the (antiquark)quark.

Equation~(\ref{EVEquation}) is a nonlinear eigenvalue equation
that cannot be solved with standard techniques. In order to study
it in some more detail we use a simple toy model in  which spin
and flavor of the (anti)quark are neglected and a real scalar
particle is taken for the Goldstone boson. We use a harmonic
oscillator confinement in the square of the mass operator. This
model has 5 parameters: the (anti)quark mass $m_q$, the
Goldstone-boson mass $m_{GB}$, the Goldstone-boson-quark coupling
strength $g$, the oscillator parameter $a$ and a parameter $V_0$
to shift the mass spectrum. We have taken a standard value of
$0.34$~GeV for $m_q$ and the pion mass for $m_{GB}$. To give our
toy model some physical meaning the parameters $a$ and $V_0$ have
been fixed in such a way that the the experimental masses of the
$\omega$ ground state and its first excited state are
approximately reproduced. The Goldstone-boson-quark coupling is
varied within the range allowed by the Goldberger-Treiman
relation, i.e. $0.67\lesssim {g^2}/{4\pi} \lesssim
1.19$~\cite{Glozman:1997fs}. To simplify things further only
radial excitations of bare mesons have been taken into account.
The mass eigenvalue problem, Eq.~(\ref{EVEquation}), can be solved
by an iterative procedure. One first has to restrict the number of
bare states, that are taken into account, to a certain number
$\alpha_{\mathrm{max}}$. The first step is to insert a start value
for $m$ into $\hat{V}_{\mathrm{opt}}(m)$ and solve the resulting
linear eigenvalue equation. This leads to $\alpha_{\mathrm{max}}$
(possibly complex) eigenvalues. From these one has to pick out the
right one, reinsert it into $\hat{V}_{\mathrm{opt}}(m)$, solve
again, etc. Appropriate start values are, e.g., the eigenvalues of
the pure confinement problem. Note that the optical potential
$V^{\alpha \alpha^{\prime}}_{\mathrm{opt}}(m)$ becomes complex if
the mass eigenvalue $m$ is larger than the lowest threshold
$m_{\mathrm{th}}=m_0+m_{GB}$, i.e. the mass of the lightest bare
meson plus the Goldstone-boson mass. As a consequence also the
physical mass eigenvalues $m$ will become complex as soon as their
real part is larger than $m_{\mathrm{th}}$ and we will get
unstable meson excitations. The mass of such an excitation can
then be identified with $\mathrm{Re}(m)$, its width $\Gamma$ with
$2\, \mathrm{Im}(m)$.
\begin{figure}
 \centering
 \includegraphics[width=12cm]{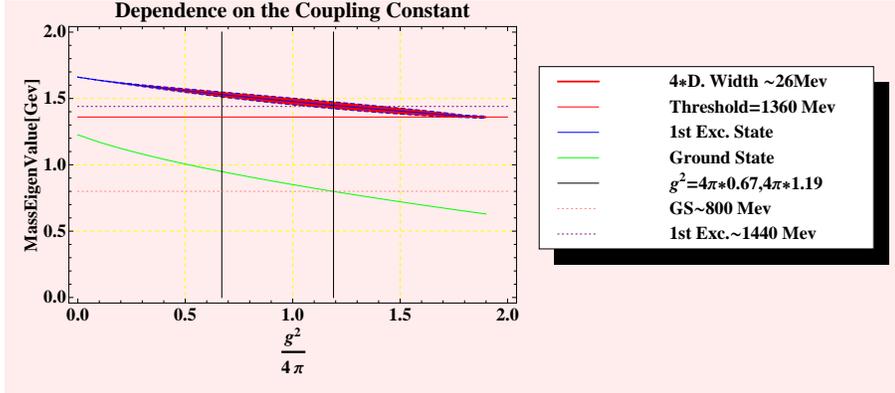}
\caption{\small{Predictions of our toy model for the meson masses
and widths. Shown are the ground state (green line) and the first
excited state (blue line) as functions of the
Goldstone-boson-quark coupling. The red band between the dashed
blue lines represents four times the decay width of the first
excited state.}} \label{DepCoCo}
 \end{figure}

The results of a first numerical study with our toy model (with
$\alpha_{\mathrm{max}}=2$) are shown in Fig.~\ref{DepCoCo}. It can
be seen that the Goldstone-boson loop provides an attractive force
and that the decay width exhibits a maximum as a function of
${g^2}/{4\pi}$. As soon as the real part of the mass eigenvalue of
the first excited state approaches $m_0+m_{\mathrm{GB}}$, where
$m_0$ is the harmonic oscillator ground-state mass, the decay
width vanishes. With a Goldstone-boson-quark coupling of
$g^2/4\pi=1.19$, which is still compatible with the
Goldberger-Treiman relation, the 2 lowest lying states are found
to have masses of about $0.8$ and $1.44$~GeV, respectively. The
first excited state has a width of $0.026$~GeV. An increase of
$\alpha_{\mathrm{max}}$ changes these values by only a few
percent.The iterative procedure converges already after 5
iterations.

These are promising results in view of the simplicity of our toy
model and it will be interesting to see whether typical decay
widths of $0.1$~GeV or more can be achieved within our approach in
the more interesting case of baryon resonances for the full chiral
constituent-quark model.

\end{document}